\documentclass[%
 reprint,
superscriptaddress,
 amsmath,amssymb,
pre,
]{revtex4-2}

\usepackage{xcolor}
\usepackage{graphicx}% Include figure files
\usepackage{dcolumn}% Align table columns on decimal point
\usepackage{bm}% bold math
\begin{document}

% \preprint{APS/PRL}

\title{Filamentation of a Relativistic Proton Bunch in Plasma }% Force line breaks with \\

\author{L.~Verra}
\email{{livio.verra@lnf.infn.it}}
\altaffiliation{Present Address: INFN Laboratori Nazionali di Frascati, 00044 Frascati, Italy}
\affiliation{CERN, 1211 Geneva 23, Switzerland}
\author{C.~Amoedo}
\affiliation{CERN, 1211 Geneva 23, Switzerland}
\author{N.~Torrado}
\affiliation{CERN, 1211 Geneva 23, Switzerland}
\affiliation{GoLP/Instituto de Plasmas e Fus\~{a}o Nuclear, Instituto Superior T\'{e}cnico, Universidade de Lisboa, 1049-001 Lisbon, Portugal}
\author{A.~Clairembaud}
\affiliation{CERN, 1211 Geneva 23, Switzerland}
\author{J.~Mezger}
\affiliation{Max Planck Institute for Physics, 80805 Munich, Germany}
\author{F.~Pannell}
\affiliation{UCL, London WC1 6BT, United Kingdom}
\author{J.~Pucek}
\affiliation{Max Planck Institute for Physics, 80805 Munich, Germany}
\author{N.~van Gils}
\affiliation{CERN, 1211 Geneva 23, Switzerland}
\author{M.~Bergamaschi}
\affiliation{Max Planck Institute for Physics, 80805 Munich, Germany}
\author{G.~Zevi Della Porta}
\affiliation{CERN, 1211 Geneva 23, Switzerland}
\affiliation{Max Planck Institute for Physics, 80805 Munich, Germany}
\author{N.~Lopes}
\affiliation{GoLP/Instituto de Plasmas e Fus\~{a}o Nuclear, Instituto Superior T\'{e}cnico, Universidade de Lisboa, 1049-001 Lisbon, Portugal}
\author{A.~Sublet}
\affiliation{CERN, 1211 Geneva 23, Switzerland}
\author{M.~Turner}
\affiliation{CERN, 1211 Geneva 23, Switzerland}
\author{E.~Gschwendtner}
\affiliation{CERN, 1211 Geneva 23, Switzerland}
\author{P.~Muggli}
\affiliation{Max Planck Institute for Physics, 80805 Munich, Germany}
\collaboration{AWAKE Collaboration}
\noaffiliation
\author{R.~Agnello}
\affiliation{Ecole Polytechnique Federale de Lausanne (EPFL), Swiss Plasma Center (SPC), 1015 Lausanne, Switzerland}
\author{C.C.~Ahdida}
\affiliation{CERN, 1211 Geneva 23, Switzerland}
\author{Y.~Andrebe}
\affiliation{Ecole Polytechnique Federale de Lausanne (EPFL), Swiss Plasma Center (SPC), 1015 Lausanne, Switzerland}
\author{O.~Apsimon}
\affiliation{University of Manchester M13 9PL, Manchester M13 9PL, United Kingdom}
\affiliation{Cockcroft Institute, Warrington WA4 4AD, United Kingdom}
\author{R.~Apsimon}
\affiliation{Cockcroft Institute, Warrington WA4 4AD, United Kingdom} % left as in previous papers
\affiliation{Lancaster University, Lancaster LA1 4YB, United Kingdom}
\author{J.M.~Arnesano}
\affiliation{CERN, 1211 Geneva 23, Switzerland}
\author{V.~Bencini}
\affiliation{CERN, 1211 Geneva 23, Switzerland}
\affiliation{John Adams Institute, Oxford University, Oxford OX1 3RH, United Kingdom}
\author{P.~Blanchard}
\affiliation{Ecole Polytechnique Federale de Lausanne (EPFL), Swiss Plasma Center (SPC), 1015 Lausanne, Switzerland}
\author{P.N.~Burrows}
\affiliation{John Adams Institute, Oxford University, Oxford OX1 3RH, United Kingdom}
\author{B.~Buttensch{\"o}n}
\affiliation{Max Planck Institute for Plasma Physics, 17491 Greifswald, Germany}
\author{A.~Caldwell}
\affiliation{Max Planck Institute for Physics, 80805 Munich, Germany}
\author{M.~Chung}
\affiliation{UNIST, Ulsan 44919, Republic of Korea}
\author{D.A.~Cooke}
\affiliation{UCL, London WC1 6BT, United Kingdom}
\author{C.~Davut}
\affiliation{University of Manchester M13 9PL, Manchester M13 9PL, United Kingdom}
\affiliation{Cockcroft Institute, Warrington WA4 4AD, United Kingdom} % left as in previous papers
\author{G.~Demeter}
\affiliation{HUN-REN Wigner Research Centre for Physics, Budapest, Hungary}
\author{A.C.~Dexter}
\affiliation{Cockcroft Institute, Warrington WA4 4AD, United Kingdom} % left as in previous papers
\affiliation{Lancaster University, Lancaster LA1 4YB, United Kingdom}
\author{S.~Doebert}
\affiliation{CERN, 1211 Geneva 23, Switzerland}
\author{J.~Farmer}
\affiliation{Max Planck Institute for Physics, 80805 Munich, Germany}
\author{A.~Fasoli}
\affiliation{Ecole Polytechnique Federale de Lausanne (EPFL), Swiss Plasma Center (SPC), 1015 Lausanne, Switzerland}
\author{R.~Fonseca}
\affiliation{ISCTE - Instituto Universit\'{e}ario de Lisboa, 1049-001 Lisbon, Portugal}  % left as in previous papers
\affiliation{GoLP/Instituto de Plasmas e Fus\~{a}o Nuclear, Instituto Superior T\'{e}cnico, Universidade de Lisboa, 1049-001 Lisbon, Portugal}
\author{I.~Furno}
\affiliation{Ecole Polytechnique Federale de Lausanne (EPFL), Swiss Plasma Center (SPC), 1015 Lausanne, Switzerland}
\author{E.~Granados}
\affiliation{CERN, 1211 Geneva 23, Switzerland}
\author{M.~Granetzny}
\affiliation{University of Wisconsin, Madison, WI 53706, USA}
\author{T.~Graubner}
\affiliation{Philipps-Universit{\"a}t Marburg, 35032 Marburg, Germany}
\author{O.~Grulke}
\affiliation{Max Planck Institute for Plasma Physics, 17491 Greifswald, Germany}
\affiliation{Technical University of Denmark, 2800 Kgs. Lyngby, Denmark}
\author{E.~Guran}
\affiliation{CERN, 1211 Geneva 23, Switzerland}
\author{J.~Henderson}
\affiliation{Cockcroft Institute, Warrington WA4 4AD, United Kingdom}
\affiliation{STFC/ASTeC, Daresbury Laboratory, Warrington WA4 4AD, United Kingdom}
\author{M.Á.~Kedves}
\affiliation{HUN-REN Wigner Research Centre for Physics, Budapest, Hungary}
\author{F.~Kraus}
\affiliation{Philipps-Universit{\"a}t Marburg, 35032 Marburg, Germany}
\author{M.~Krupa}
\affiliation{CERN, 1211 Geneva 23, Switzerland}
\author{T.~Lefevre}
\affiliation{CERN, 1211 Geneva 23, Switzerland}
\author{L.~Liang}
\affiliation{University of Manchester M13 9PL, Manchester M13 9PL, United Kingdom}
\affiliation{Cockcroft Institute, Warrington WA4 4AD, United Kingdom}
\author{S.~Liu}
\affiliation{TRIUMF, Vancouver, BC V6T 2A3, Canada}
\author{K.~Lotov}
\affiliation{Budker Institute of Nuclear Physics SB RAS, 630090 Novosibirsk, Russia}
\affiliation{Novosibirsk State University, 630090 Novosibirsk , Russia}
\author{M.~Martinez~Calderon}
\affiliation{CERN, 1211 Geneva 23, Switzerland}
\author{S.~Mazzoni}
\affiliation{CERN, 1211 Geneva 23, Switzerland}
\author{K.~Moon}
\affiliation{UNIST, Ulsan 44919, Republic of Korea}
\author{P.I.~Morales~Guzm\'{a}n}
\affiliation{Max Planck Institute for Physics, 80805 Munich, Germany}
\author{M.~Moreira}
\affiliation{GoLP/Instituto de Plasmas e Fus\~{a}o Nuclear, Instituto Superior T\'{e}cnico, Universidade de Lisboa, 1049-001 Lisbon, Portugal}
\author{T.~Nechaeva}
\affiliation{Max Planck Institute for Physics, 80805 Munich, Germany}
\author{N.~Okhotnikov}
\affiliation{Budker Institute of Nuclear Physics SB RAS, 630090 Novosibirsk, Russia}
\affiliation{Novosibirsk State University, 630090 Novosibirsk , Russia}
\author{C.~Pakuza}
\affiliation{John Adams Institute, Oxford University, Oxford OX1 3RH, United Kingdom}
\author{A.~Pardons}
\affiliation{CERN, 1211 Geneva 23, Switzerland}
\author{K.~Pepitone}
\affiliation{Angstrom Laboratory, Department of Physics and Astronomy, 752 37 Uppsala, Sweden}
\author{E.~Poimendidou}
\affiliation{CERN, 1211 Geneva 23, Switzerland}
\author{A.~Pukhov}
\affiliation{Heinrich-Heine-Universit{\"a}t D{\"u}sseldorf, 40225 D{\"u}sseldorf, Germany}
\author{R.L.~Ramjiawan}
\affiliation{CERN, 1211 Geneva 23, Switzerland}
\affiliation{John Adams Institute, Oxford University, Oxford OX1 3RH, United Kingdom}
\author{L.~Ranc}
\affiliation{Max Planck Institute for Physics, 80805 Munich, Germany}
\author{S.~Rey}
\affiliation{CERN, 1211 Geneva 23, Switzerland}
\author{R.~Rossel}
\affiliation{CERN, 1211 Geneva 23, Switzerland}
\author{H.~Saberi}
\affiliation{University of Manchester M13 9PL, Manchester M13 9PL, United Kingdom}
\affiliation{Cockcroft Institute, Warrington WA4 4AD, United Kingdom}
\author{O.~Schmitz}
\affiliation{University of Wisconsin, Madison, WI 53706, USA}
\author{E.~Senes}
\affiliation{CERN, 1211 Geneva 23, Switzerland}
\author{F.~Silva}
\affiliation{INESC-ID, Instituto Superior Técnico, Universidade de Lisboa, 1049-001 Lisbon, Portugal}
\author{L.~Silva}
\affiliation{GoLP/Instituto de Plasmas e Fus\~{a}o Nuclear, Instituto Superior T\'{e}cnico, Universidade de Lisboa, 1049-001 Lisbon, Portugal}
\author{B.~Spear}
\affiliation{John Adams Institute, Oxford University, Oxford OX1 3RH, United Kingdom}
\author{C.~Stollberg}
\affiliation{Ecole Polytechnique Federale de Lausanne (EPFL), Swiss Plasma Center (SPC), 1015 Lausanne, Switzerland}
\author{C.~Swain}
\affiliation{Cockcroft Institute, Warrington WA4 4AD, United Kingdom}
\affiliation{University of Liverpool, Liverpool L69 7ZE, United Kingdom}
\author{A.~Topaloudis}
\affiliation{CERN, 1211 Geneva 23, Switzerland}
\author{P.~Tuev}
\affiliation{Budker Institute of Nuclear Physics SB RAS, 630090 Novosibirsk, Russia}
\affiliation{Novosibirsk State University, 630090 Novosibirsk , Russia}
\author{F.~Velotti}
\affiliation{CERN, 1211 Geneva 23, Switzerland}
\author{V.~Verzilov}
\affiliation{TRIUMF, Vancouver, BC V6T 2A3, Canada}
\author{J.~Vieira}
\affiliation{GoLP/Instituto de Plasmas e Fus\~{a}o Nuclear, Instituto Superior T\'{e}cnico, Universidade de Lisboa, 1049-001 Lisbon, Portugal}
\author{E.~Walter}
\affiliation{Max Planck Institute for Plasma Physics, Munich, Germany}
\author{C.~Welsch}
\affiliation{Cockcroft Institute, Warrington WA4 4AD, United Kingdom}
\affiliation{University of Liverpool, Liverpool L69 7ZE, United Kingdom}
\author{M.~Wendt}
\affiliation{CERN, 1211 Geneva 23, Switzerland}
\author{M.~Wing}
\affiliation{UCL, London WC1 6BT, United Kingdom}
\author{J.~Wolfenden}
\affiliation{Cockcroft Institute, Warrington WA4 4AD, United Kingdom}
\affiliation{University of Liverpool, Liverpool L69 7ZE, United Kingdom}
\author{B.~Woolley}
\affiliation{CERN, 1211 Geneva 23, Switzerland}
\author{G.~Xia}
\affiliation{Cockcroft Institute, Warrington WA4 4AD, United Kingdom}
\affiliation{University of Manchester M13 9PL, Manchester M13 9PL, United Kingdom}
\author{V.~Yarygova}
\affiliation{Budker Institute of Nuclear Physics SB RAS, 630090 Novosibirsk, Russia}
\affiliation{Novosibirsk State University, 630090 Novosibirsk , Russia}
\author{M.~Zepp}
\affiliation{University of Wisconsin, Madison, WI 53706, USA}
\noaffiliation
% \author{Collaboration List}
% \noaffiliation

\date{\today}% It is always \today, today,
%              %  but any date may be explicitly specified

\begin{abstract}
% %%abstract 1%%
% Abstract 1: Wide charged particle bunches propagating in plasma undergo the current filamentation instability. 
% Using a long, relativistic proton bunch we experimentally demonstrate the occurrence of the instability, \textcolor{red}{in the form of the oblique instability}, and we determine a threshold value for the ratio between the bunch transverse size and plasma skin depth for the instability to occur. 
% \textcolor{red}{We observe that, at the threshold, the outcome of the experiment alternates between filamentation and self-modulation instability.}
% Time-resolved images of the bunch density reveal that filamentation grows to an observable level late along the bunch, confirming the spatio-temporal nature of the instability. 
% We calculate the amplitude of the magnetic field generated in the plasma and show that the associated magnetic energy increases with plasma density.

% \textcolor{red}{Abstract 2: }
% The combination of the transverse two-stream and current filamentation instabilities, known as oblique instability, occurs when wide, long and underdense relativistic charged particle bunches propagate in plasma. 

We show in experiments that a long, underdense, relativistic proton bunch propagating in plasma undergoes the oblique instability, that we observe as filamentation. 
We determine a threshold value for the ratio between the bunch transverse size and plasma skin depth for the instability to occur. 
At the threshold, the outcome of the experiment alternates between filamentation and self-modulation instability (evidenced by longitudinal modulation into microbunches).
Time-resolved images of the bunch density distribution reveal that filamentation grows to an observable level late along the bunch, confirming the spatio-temporal nature of the instability. 
We calculate the amplitude of the magnetic field generated in the plasma by the instability and show that the associated magnetic energy increases with plasma density.

\end{abstract}

\maketitle
\section{Introduction}
When a perturbation is introduced in a plasma, e.g. by a streaming relativistic, charged particle bunch, plasma electrons move to restore the charge and current neutrality of the system. 
Compensation of the space-charge field of the bunch can cause oscillatory phenomena such as plasma wakefields~\cite{PWFA:CHEN}.
Depending on the bunch spatial dimensions, compared to the physical quantities (e.g. plasma wavenumber and skin depth) determining the unstable modes, the bunch-plasma system can be prone to different instability regimes~\cite{instabilities,BRET:REVIEW}.

\par When the root mean square (rms) length of the bunch $\sigma_z$ is much longer than the plasma skin depth $\delta = c/\omega_{pe}$, where $c$ is the speed of light and $\omega_{pe}=\sqrt{n_{pe}e^2/m_e \varepsilon_0}$ is the plasma electron angular frequency ($n_{pe}$ the plasma electron density, $e$ the elementary charge, $m_e$ the electron mass at rest and $\varepsilon_0$ the vacuum permittivity),
wakefields act on the bunch itself. 
When the bunch rms transverse size $\sigma_{r0}\lesssim \delta$, the axisymmetric mode of the transverse two-stream instability (TTSI)~\cite{PIERCE:10.1063/1.1715050,BOHM:PhysRev.75.1851}, i.e., the self-modulation instability (SMI)~\cite{KUMAR:GROWTH,KARL:PRL,MARLENE:PRL,LIVIO:WIDE}, can take place.
SMI modulates longitudinally the density of the bunch along its axis with wavenumber $k = k_{\parallel} = \delta^{-1}$. %
The resulting microbunch train then resonantly drives large-amplitude wakefields that can be used for particle acceleration, as in the AWAKE experiment at CERN~\cite{PATRIC:READINESS,KARL:PRL,MARLENE:PRL,AW:NATURE}.

\par When $\sigma_{r0}>\delta$, the plasma return current can flow within the bunch.
When the peak bunch to plasma electron density ratio $n_{b0}/ n_{pe}\sim 1$ and the Lorentz factor of the bunch $\gamma\sim 1$, repulsion between opposite currents tends to reinforce any transverse anisotropy in the current density distributions, possibly leading to the development of the current filamentation instability (CFI)~\cite{LEE:PhysRevLett.31.1390%
,BRET:REVIEW}. 
This instability modulates the bunch transversely ($k=k_{\perp}$) into multiple filaments, that self-pinch to transverse size on the order of $\delta$~\cite{LEE:PhysRevLett.31.1390} and with a higher current density than that of the incoming bunch. %

\par Calculations and numerical simulation results show that a wide spectrum of wavenumbers is excited~\cite{CFI_Silva_2003,LEE:PhysRevLett.31.1390,CALIFANO,GEDALIN}.
Occurrence of CFI generates magnetic fields within the medium, thus converting part of the kinetic energy carried by streaming particles into magnetic energy ~\cite{LU:CFI,HUNTINGTON:CFI,PhysRevLett.131.055201}.
This process is in fact one of the plausible candidates for magnetization of astrophysical media~\cite{Magnetization_Schlickeiser_2003,Unmagnetized_Medvedev_2006}, as well as for the magnetic fields enhancement that could explain phenomena such as long-duration afterglow of gamma-ray bursts~\cite{Medvedev_2007,Medvedev_2009} and collisionless shocks~\cite{Medvedev_1999}.
It also plays an important role in the transport and deposition of energy by hot electrons in inertial confinement fusion targets~\cite{Fusion_PhysRevLett.85.2128,Fusion_10.1063/1.1476004}.
Previous experiments using electron bunches~\cite{TARAKIS:CFI,HUNTINGTON:CFI_PRL,CFI:PhysRevLett.109.185007} investigated conditions where the instability reached saturation and filaments even started merging with each other, a process firstly described in~\cite{Merging_Medvedev_2005}.
The effect of the return current on the focusing of an electron bunch was observed in a passive plasma lens with $\sigma_{r0}\gtrsim\delta$~\cite{GOVIL:PhysRevLett.83.3202}. 
CFI is an unwanted instability for effective plasma wakefield excitation, because it would degrade the structure of the wakefields and therefore the emittance of the accelerated bunch. 

\par Theory~\cite{BRET:REVIEW,CALIFANO:PhysRevE.58.7837, Bret_2009} and numerical simulation results~\cite{shukla_vieira_muggli_sarri_fonseca_silva_2018, OTSI:PhysRevResearch.4.023085} show that in the case of relativistic ($\gamma\gg 1$), wide, long and underdense ($n_{b0}/ n_{pe}\ll 1$) bunches, the oblique instability (OBI) is the dominant process. %
In this case, the wavenumber retains perpendicular and parallel components: $\vec{k}=\vec{k}_{\perp}+\vec{k}_{\parallel}$.
OBI thus tends to generate finite-length filaments surrounded by the return current of plasma electrons moving non-relativistically with speed $v_e\sim (n_{b0}/n_{pe})c\ll c$.
In the following we thus refer to this phenomenon with the general term $\textit{filamentation}$.

\par Using a long, relativistic  proton ($p^+$) bunch, we perform experiments, in the context of AWAKE, in the very underdense ($n_{b0}/n_{pe}\le10^{-2}$) and very relativistic ($\gamma_p$=427) regime. %
Protons have mass $m_p~\sim 1836\, m_e$, increasing the inertia ($\gamma_p m_p$) of the bunch particles and slowing the growth of instabilities, when compared to the most described case of an electron bunch with $\gamma<160$. %
OBI is therefore expected to be the dominant instability. %
Both bunch (relative energy spread~$\sim0.2\%$) and plasma (electron temperature~$\sim$few\,eV) can be considered as cold. %
The growth rate of OBI is given by:
\begin{equation}
    \label{eq:1}
    \Gamma = \Gamma_e \sqrt{\frac{m_e}{m_p}}= \frac{\sqrt{3}}{2^{4/3}} ({\frac{n_{b0}}{n_{pe}\gamma_p}})^{1/3} \omega_{pe} \sqrt{\frac{m_e}{m_p}, }
\end{equation}
where $\Gamma_e$ is the growth rate for the case of an electron bunch~\cite{BRET:REVIEW}.

\par In this paper, we show with experimental results that, when $\sigma_{r0}/\delta\geq1.5$, OBI, and therefore filamentation, occurs.
We observe the presence of filaments in transverse time-integrated images of the bunch after propagation through 10\,m of plasma. 
Time-resolved images of a transverse slice of the bunch indicate that filaments become observable only late along the bunch, consistent with
small growth rate and the early stage of OBI, and with the spatio-temporal nature of filamentation of a finite length bunch~\cite{Growth_Pathak_2015,SLAC:PhysRevResearch.2.023123,Shukla_2020, shukla_vieira_muggli_sarri_fonseca_silva_2018, OTSI:PhysRevResearch.4.023085}.
We observe that for $\sigma_{r0}/\delta=1.5$, and without observable differences in the incoming time-integrated bunch current density distribution, the outcome of the experiment alternates between filamentation (evidenced by transverse modulation) and SMI (evidenced by longitudinal modulation). 
For values below this threshold, SMI always takes place and filamentation is not observed from time-integrated images of the bunch. 
We estimate the amplitude of the magnetic field generated by the occurrence of filamentation.
We calculate that the amount of magnetic energy converted from kinetic energy is small but increases with the plasma electron density, due to an increase of the number of filaments in a larger area of the bunch.

\section{Experimental Setup}

\par
\begin{figure}[h!]
\centering
\includegraphics[width=\columnwidth]{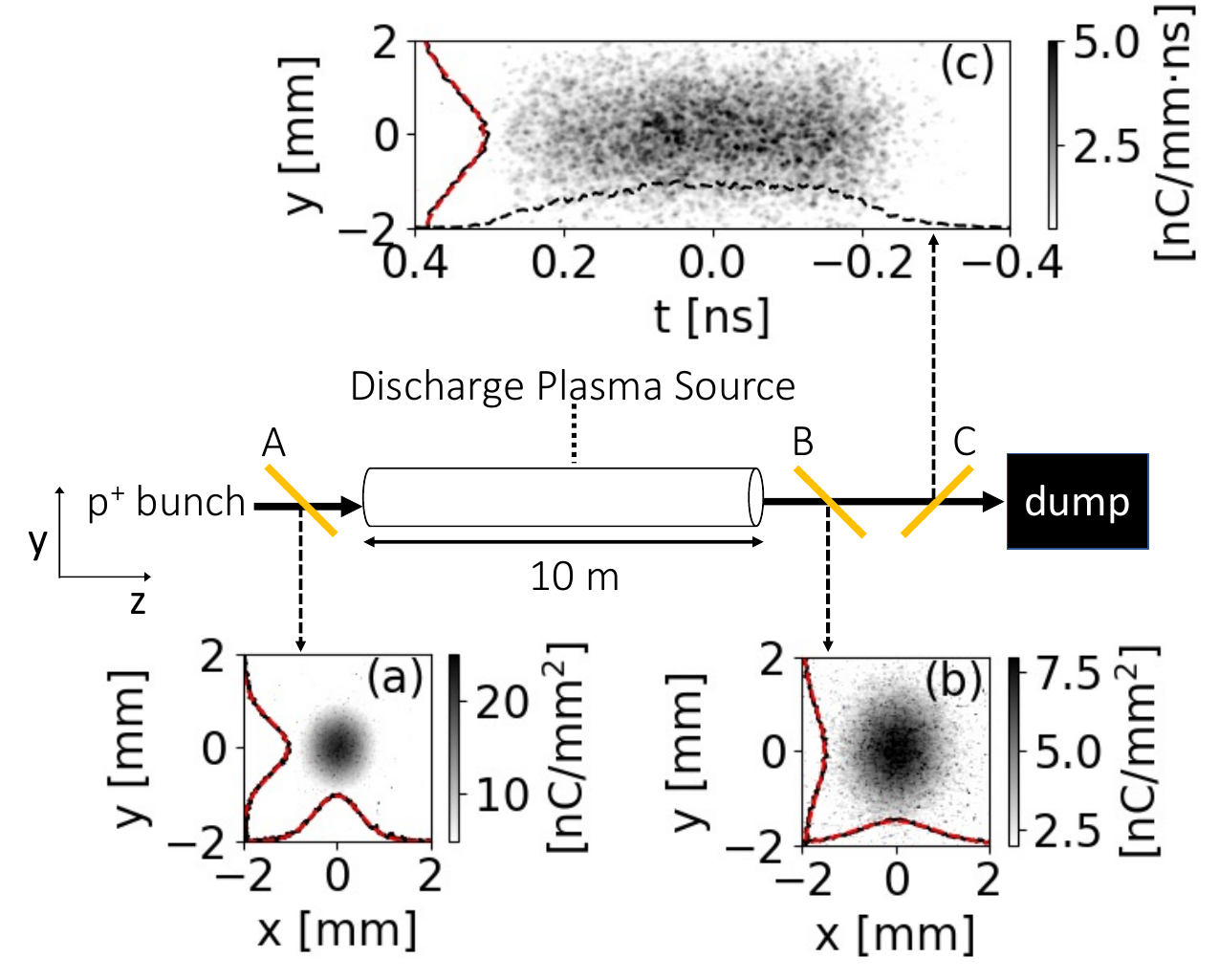}
\caption{
Schematic of the AWAKE experimental setup (not to scale).
Insets: typical images of the bunch with neither plasma nor gas in the source.
(a,b): transverse time-integrated images of the bunch at screens A and B, respectively.
(c): time-resolved image at screen C.
Solid black lines in the insets: transverse projections of the distributions.
Red dashed lines: Gaussian fits of the transverse projections. 
Dashed black line in inset (c): longitudinal projection. 
}
\label{fig:1}
\end{figure}

 The plasma (see Fig.~\ref{fig:1}) is generated by a DC-pulsed discharge in a 10-m-long, 25-mm-diameter, cylindrical glass tube filled with argon at pressure $23.8$\,Pa~\cite{TORRADO:DISCHARGE}.
The plasma fills the entire diameter of the tube.
The peak discharge current used for these experiments is $\sim500\,$A, with a pulse duration of $\sim 25\,$\textmu s.
%%%%%%%%%%%%%%%%%%%
%%%%%%%%%%%%%%%%%%%%%%%%
\par The 43\,nC bunch with normalized emittance $\epsilon_N = 2.5$\,mm-mrad composed of 400\,GeV/c $p^+$ is delivered by the CERN Super Proton Synchrotron. 
By varying the delay between the $p^+$ bunch arrival time and the start of the discharge between 25 and 200\,\textmu s, we vary the plasma electron density between $n_{pe}=9.38$ and $0.68\times 10^{14}\,$cm$^{-3}$%
, respectively, and thus the value of $\sigma_{r0}/\delta$ ($\sigma_{r0}$ constant), due to recombination of the plasma. 
The values of $n_{pe}$ as a function of delay were measured offline by longitudinal double-pass interferometry~\cite{CAROLINA:DISCHARGE} and, in these experiments, from the frequency of the microbunch train resulting from SMI, with bunches with small transverse size ($\sigma_{r0}\sim0.2\,$mm) such that $\sigma_{r0}/\delta< 1$~\cite{KARL:PRL}.

\par We measure the bunch current density distribution using aluminum-coated silicon wafers placed at $45^\circ$ in the beam path, emitting optical transition radiation (OTR) when protons enter them.
Backward OTR is imaged onto the chips of CMOS cameras (screens A and B, Fig.~\ref{fig:1}) or onto the entrance slit of a streak camera (screen C), to obtain, respectively, transverse time-integrated or time-resolved images of the bunch.
Insets (a,b) and (c) show the corresponding images of the $p^+$ bunch while neither gas, nor discharge are present in the source. 

\par For the measurements presented here we employ beam optics providing a bunch transverse size larger than that normally used for wakefield acceleration and SMI experiments (i.e., $\sigma_{r0}> 0.2\,$mm)~\cite{LIVIO:WIDE}.
We perform Gaussian fits (red dashed lines) on the transverse projections (black solid lines) to determine the rms transverse size $\sigma_{x,y}$ of the bunch at the various screens.
We measure $\sigma_{x,y} = (0.45, 0.54)\pm 0.01\,$mm (the uncertainty corresponds to the standard deviation of 20 consecutive events) at screen A, $\sim1.3\,$m upstream of the plasma entrance %
($\sigma_{r0}$=0.50\,mm).
Since the distance from the plasma entrance is shorter than the Twiss parameter of the $p^+$ beam $\beta_0\sim 29\,$m, these images provide a good estimate of the transverse current density distribution and size of the bunch entering the source, for every event, with and without plasma. 
As the difference between sizes in the two planes is small and not relevant for the measurements described here, in the following we use the average of the two values to calculate the ratio $\sigma_{r0}/\delta$.

\par At screen B, the transverse sizes %
obtained with the same procedure are $\sigma_{x,y} = (0.80, 0.87)\pm 0.02\,$mm %
($\sigma_r$=0.84\,mm).
%%%%%%%%%%%%%%%%%%%
\begin{figure*}[hbt!]
\centering
\includegraphics[width=\textwidth]{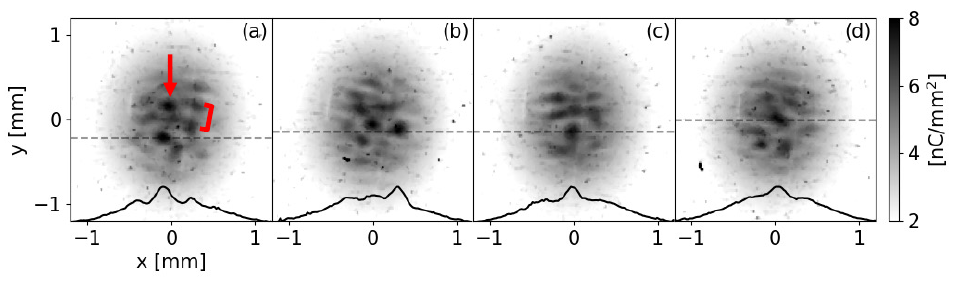}
\caption{(a-d): transverse, time-integrated, single-event images of the $p^+$ bunch after propagation in plasma at screen B (see Fig.~\ref{fig:1}) with $n_{pe} = 9.38\times 10^{14}\,$cm$^{-3}$ and $\sigma_{r0}/\delta = 2.9$.
Black lines: lineouts along the corresponding dashed grey lines.
Colormap and lineouts chosen to enhance visibility of the filaments ($\geq2\,$nC/mm$^2$).
}
\label{fig:2}
\end{figure*}
%%%%%%%%%%%%%%%%%%%%%%%%
This screen is purposely placed as close as possible to the plasma exit ($\sim0.3$\,m$\ll \beta_0$ downstream of it), because we expect filaments with size a fraction of that of the bunch, and likely higher emittance, to diverge strongly while traveling in vacuum from the plasma exit to the screens.
Their visibility is therefore expected to decrease with propagation distance.
The spatial resolution at screen B is $\sim0.027\,$mm, measured as the
50\% modulation of the transfer function.
This is much smaller than the expected transverse size of the filaments: for the maximum density used in these experiments $n_{pe}\leq9.38\times 10^{14}\,$cm$^{-3}$, $\delta \geq 0.17\,$mm.
We note here that images of the bunch before and after the source are smooth, i.e. they show no features at the expected scale size of the filaments ($\sim \delta$).

\par Screen C is positioned $\sim 3.5\,$m downstream of the plasma exit. 
Thus, narrow features of the bunch (i.e., filaments) leaving the plasma with large divergence may not be as clearly recognizable on time-resolved images as on time-integrated ones.
We measure the bunch duration $\sigma_t = (163\pm 3)\,$ps as the rms of the longitudinal projection of time-resolved images with no gas in the source (black dashed line on Fig.~\ref{fig:1}(c)).
Therefore, $\sigma_z = \sigma_t c\gg \delta$ and $n_{b0}=1.3\cdot 10^{12}\,$cm$^{-3}< 10^{-2}\, n_{pe}$ at the plasma entrance for all measurements presented here.

\section{Experimental Results}

\par Figures~\ref{fig:2}(a-d) show four, single-event, time-integrated images of the $p^+$ bunch at the screen close to the plasma exit (B, Fig.~\ref{fig:1}) after propagation in plasma with $n_{pe}=9.38\times 10^{14}\,$cm$^{-3}$, thus $\sigma_{r0}/\delta = 2.9$.
All images show clear evidence of filaments within a radius in the bunch $r\sim0.5\,$mm$<\sigma_r$, while the overall Gaussian distribution of the bunch is maintained, as indicated by the horizontal lineouts (black lines, lineouts chosen along the dashed grey lines to evidence the presence of filaments).
This indicates that, beyond a given radius of the bunch, the charge density and growth rate are not large enough for filamentation to develop over 10\,m of plasma.
The rms size of a typical filament such as the one indicated by the red arrow in Fig.~\ref{fig:2}(a) is $\sim0.12\,$mm, and the distance between neighbouring filaments as those indicated by the red bracket is $\sim0.27\,
$mm.
These values are commensurate with $\delta=0.17\,$mm, as expected from filamentation (OBI or CFI)~\cite{CFI:PhysRevLett.109.185007}.
The spaces between filaments correspond to regions where plasma return current flows~\cite{LEE:PhysRevLett.31.1390}.
A hollow ring of bunch charge centered on a filament is for example clearly visible in (c).

\par Because the transverse component of the instability grows from non-uniformities in the transverse distribution of the bunch  and return current densities ($k=k_{\perp}$), the location, shape and number of the filaments change from event to event with no recognizable pattern in the incoming distribution (monitored with screen A, images not shown, but similar to Fig.~\ref{fig:1}(a)).
The instability took place on all 64 events of the dataset, as evidenced by clear signs of filaments on every image (see Supplemental Material~\cite{suppl}).

\par Figure~\ref{fig:3} shows the time-resolved image of the entire $p^+$ bunch (traveling from left to right), obtained at screen C, corresponding to the event of Fig.~\ref{fig:2}(b). 
For $t<0.2\,$ns (i.e., over most of the bunch duration), the transverse distribution remains essentially Gaussian as in the case with no plasma (dashed black line shows the transverse projection around the grey dashed line, for $-0.1<t<-0.06\,$ns), with no sign of occurrence of SMI or OBI (filaments).
In case of SMI, the transverse distribution would widen along the bunch, forming a halo around the microbunch train~\cite{LIVIO:PRL,LIVIO:WIDE}.
This absence of SMI is confirmed by Fourier analysis of the longitudinal bunch density distribution (see below), and it is agreement with previous results we reported~\cite{LIVIO:WIDE}, showing that, when increasing the size of the $\sigma_{r0}\sim 0.2\,$mm bunch, which always experiences SMI, to $\sigma_{r0}\sim 0.5\,$mm (as in this experiment) while keeping $n_{pe}$ constant, SMI does not develop anymore.
This occurs because increasing $\sigma_{r0}$ decreases the bunch density ($\propto \sigma_{r0}^{-2}$), and therefore also decreases the amplitude of the initial wakefields from which SMI can grow.
The resulting focusing force %
from noise wakefields becomes too small to overcome the natural divergence of the bunch due to its emittance. 

\par Even though filaments strongly diverge before reaching screen C (their relative size becomes much larger than at screen B, as seen on time-integrated images at screen C, not shown), the transverse distribution in the back of the bunch (continuous black line shows the transverse projection along the grey continuous line,  for $0.24<t<0.28\,$ns) shows evidence of splitting of the distribution within that time slice, consistent with the formation of filaments in the back of the bunch.
This late occurrence along the bunch is also consistent with the only partial density modulation observed on time-integrated images and on their transverse distributions (Fig.~\ref{fig:2}).
This confirms the spatio-temporal nature of filamentation of a bunch with finite duration, as observed in simulations~\cite{SLAC:PhysRevResearch.2.023123,Shukla_2020, OTSI:PhysRevResearch.4.023085} and shown in theory~\cite{LEE:PhysRevLett.31.1390,CALIFANO,GEDALIN}.
It is also consistent with the small number of exponentiation of OBI $\Gamma z\sim0.24$ (Eq.~\ref{eq:1}) expected from the parameters of these experiments. 
We therefore observe the result of the early stage of OBI.
We note here that time-resolved images show a less clear evidence of filamentation than the time-integrated ones (see Supplemental Material~\cite{suppl} for the entire dataset). 
This is because the streak camera only captures a transverse slice of the $p^+$ bunch. 
To obtain the best time resolution, we choose a 80-\textmu m-wide wide slice through the center of the bunch.
Thus, filaments can only be observed when they form within the direction of the slice.

\par As discussed in~\cite{LIVIO:WIDE}, decreasing $n_{pe}$, while keeping $\sigma_{r0}$ constant, increases the amplitude of the initial wakefields, because the fraction of the $p^+$ bunch charge contained within $\delta$ increases. 
Moreover, when $\sigma_{r0}/\delta$ approaches unity, the plasma return current starts flowing outside of the bunch, favoring focusing into a single ``filament"~\cite{GOVIL:PhysRevLett.83.3202} and the development of SMI over that of filamentation.
In this experiment, we consistently observe the occurrence of filamentation (transverse modulation with no longitudinal modulation) also on all events with plasma densities $n_{pe}=7.37$ and $5.82\times 10^{14}$\,cm$^{-3}$, corresponding to $\sigma_{r0}/\delta=2.5$ and 2.3.
%%%%%%%%%%%%%%%%%%%
\begin{figure}[h!]
\centering
\includegraphics[width=\columnwidth]{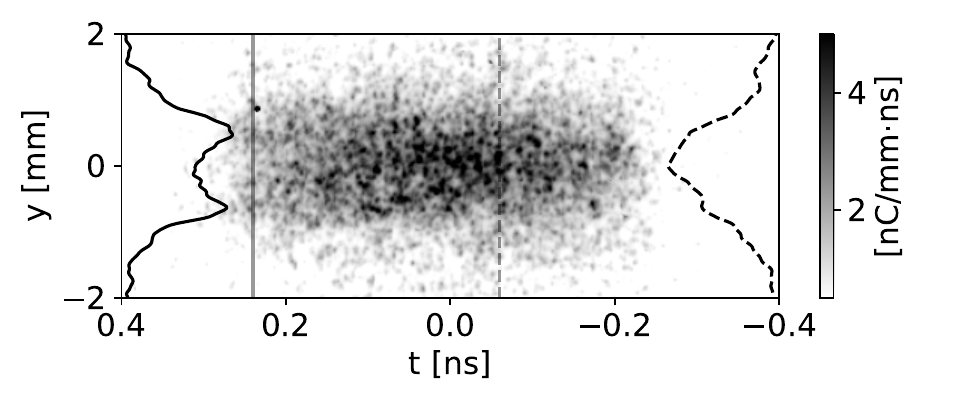}
\caption{ns-scale time-resolved, single-event image at screen C (see Fig.~\ref{fig:1}) for the event (b) of Fig.~\ref{fig:2}. 
$n_{pe} = 9.38\times 10^{14}\,$cm$^{-3}$, $\sigma_{r0}/\delta = 2.9$.
Front of the bunch at $t<0$.
Black dashed and continuous lines: transverse projection for $-0.1<t<-0.06\,$ns (front of the bunch, along the dashed grey line) and $0.24<t<0.28\,$ns (back of the bunch, along the continuous grey line), respectively.
}
\label{fig:3}
\end{figure}
%%%%%%%%%%%%%%%%%%%%%%%%
\par
\begin{figure}[h!]
\centering
\includegraphics[width=\columnwidth]{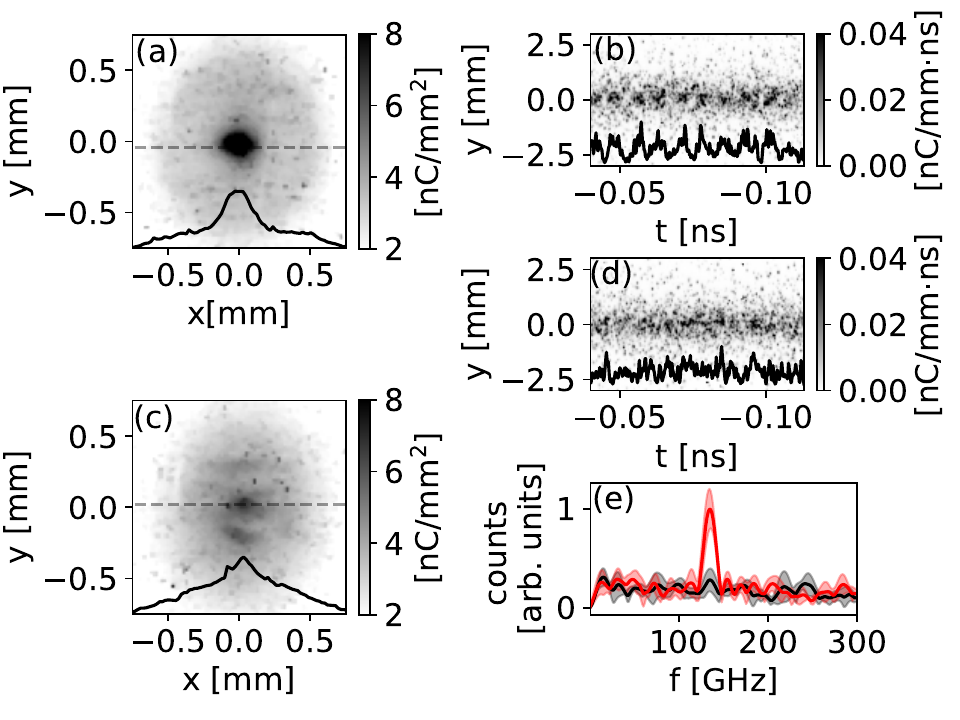}
\caption{
(a) and (c):  transverse, time-integrated, single-event images of the $p^+$ bunch at screen B (see Fig.~\ref{fig:1}) after propagation in plasma with $n_{pe} = 2.25\times10^{14}\,$cm$^{-3}$, $\sigma_{r0}/\delta=1.5$.
Black lines: lineout along the corresponding dashed grey lines.
(b) and (d): ps-scale time-resolved, single-event images at screen C, corresponding to (a) and (c), respectively.
Black lines: on-axis longitudinal distributions obtained by summing counts over $-0.217 \leq y \leq 0.217 \,$mm.
(e): average power spectra from DFT of on-axis distributions of single-event images (as in (b) and (d)).
Average amplitude of the distributions subtracted to remove the peak at $f=0$.
Red line: average of the power spectra of 20 events showing a single ``filament" on screen B (e.g., (a)); black line: average of the power spectra of 23 events showing multiple filaments on screen B (e.g., (c)).
Shaded areas show the extent of the rms variations over the events.
}
\label{fig:4}
\end{figure}

 When decreasing $n_{pe}$ further to $2.25\times 10^{14}$\,cm$^{-3}$, thus $\sigma_{r0}/\delta$ to 1.5, features on time-integrated images obtained at screen B alternate between events with one, on-axis ``filament" (Fig.~\ref{fig:4}(a)), and events with multiple filaments (Fig.~\ref{fig:4}(c)).
Over 43 consecutive events collected at this density, 20 show one on-axis ``filament" and 23 show multiple filaments, indicating an almost even alternation.

\par When only one ``filament" occurs, time-resolved images at screen C (e.g., Fig.~\ref{fig:4}(b), ps-scale) show a microbunch train %
(i.e., longitudinal modulation of the density all along the bunch), and the on-axis longitudinal distributions (black line on Fig.~\ref{fig:4}(b)) exhibit a periodic modulation, indicating that SMI has taken place. 
The average of the power spectra obtained from discrete Fourier transform (DFT) analysis of the longitudinal distributions of single-event images, for the 20 cases when the time-integrated images show only one ``filament", is shown in Fig.~\ref{fig:4}(e) (red line, the shaded area represents the rms variation).
The spectrum has a clear peak at $f_{mod}=135\pm 1\,$GHz (the uncertainty is the standard deviation of the measured values), in agreement with the value of the plasma electron frequency $f_{pe}=\omega_{pe}/2\pi=134.5\,$GHz, as typical of SMI~\cite{KARL:PRL}.
This is consistent with the transverse distribution of Fig.~\ref{fig:4}(a) (black line shows the lineout along the dashed grey line) showing bright core and wide halo, corresponding to the microbunch train and to the defocused protons, respectively~\cite{MARLENE:PRL}.

\par Conversely, the average power spectrum of the images for the 23 events with multiple filaments (black line in Fig.~\ref{fig:4}(e)) does not have any peak above the noise level, consistent with the distribution of Fig.~\ref{fig:4}(d) showing no particular periodicity, i.e. no %
detectable occurrence of longitudinal modulation (SMI or OBI) along the core of the bunch.
In this case, filaments occur within a radius of the bunch $r\sim0.25\,$mm, smaller than in the higher $n_{pe}$ and smaller $\delta$ case (Fig.~\ref{fig:2}). 
The transverse modulation also appears less deep than at higher densities.

\par These results indicate that the parameters of the experiment are such that, over the 10-m-long plasma, filamentation grows to an observable level, only late along the bunch (Fig.~\ref{fig:3}), when $\sigma_{r0}/\delta>1.5$, and no periodic modulation develops, as shown with SMI in~\cite{LIVIO:WIDE}.
When $\sigma_{r0}/\delta=1.5$, we observe multiple filaments only when SMI does not develop (Fig.~\ref{fig:4}(c,d)).
When it occurs, SMI grows to a significant level over most of the bunch (starting from its front)~\cite{LIVIO:PRL,LIVIO:WIDE} and thus dominates over the possible filamentation. 
When filamentation occurs, it develops only late along the bunch and we observe that it splits the bunch charge among short filaments, but we detect no longitudinal component of modulation. %
%%%%%%%%%%%%%%%%%%%
%%%%%%%%%%%%%%%%%%%%%%%%
\par 
The fact that we do not observe longitudinal modulation in the conditions where filamentation occurs (see Figs.~\ref{fig:3}, \ref{fig:4} and Ref.~\cite{LIVIO:WIDE}) indicates that the effect of the longitudinal component ($k_{\parallel}$) of OBI is much smaller than the effect of the transverse component ($k_{\perp}$), possibly also in agreement with the early stage of the instability.
We also note that longitudinal components of OBI recorded in e.g. two filaments on the time-resolved images would have the same frequency but likely different relative phases, which would lower the amplitude of the corresponding peak in the DFT average power spectrum, making detection more difficult. 
We therefore calculate the amplitude of the magnetic fields generated by OBI as only due to the filamentation we observe.

\par When filamentation grows, the total net current and magnetic field remain zero on average, but the formation and pinching of the filaments with higher current density than that of the bunch without filaments, and the expulsion of the return current, generate a net local magnetic field.
Since at the early stage of filamentation the overall Gaussian transverse distribution of the bunch is preserved, we use the time-integrated images from screen B (e.g., Fig.~\ref{fig:2}(a-d))  to estimate the magnetic field generated by the bunch-plasma interaction. 
We generate the distribution of the return current as the complementary of the bunch current density two-dimensional distribution with respect to the underlying smooth bunch distribution, so that the total current is zero.
The smooth distribution is obtained by applying a  median filter of size (0.38\,mm)$^2$, larger than the transverse size of the filaments, to the time-integrated images of the bunch after propagation in plasma.
We use this smoothed distribution and not the plasma off one because of the global focusing the bunch experiences along the plasma~\cite{LIVIO:AAC22}.
We then use Amp\`{e}re's law to calculate the amplitude of the transverse magnetic field due to filamentation as the sum of the fields generated by the two current densities.
To calculate the current density in each pixel of the images, we use the bunch charge density %
(see Figs.~\ref{fig:2} and \ref{fig:4}) divided by the bunch duration $\sigma_t$.
%%%%%%%%%%%%%%%%%%%
\begin{figure}[h!]
\centering
\includegraphics[width=\columnwidth]{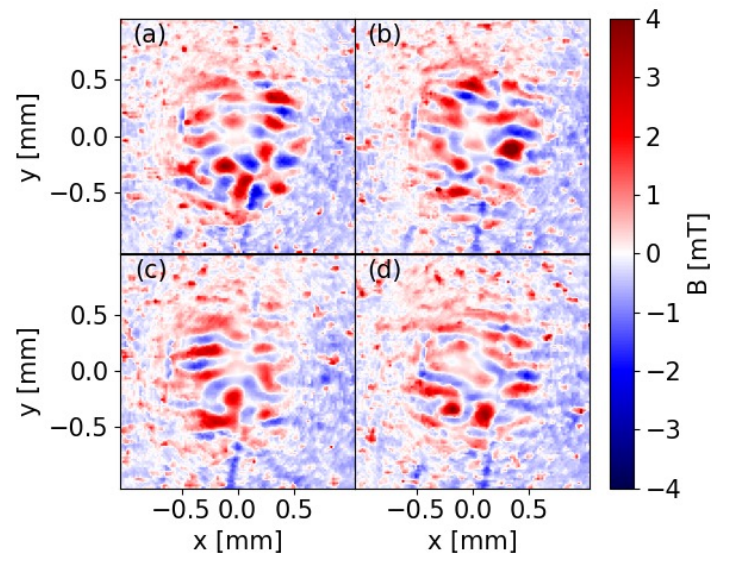}
\caption{
(a-d): %amplitude 
magnitude %
of the transverse magnetic field calculated from the images in Fig.~\ref{fig:2}(a-d).
$n_{pe} = 9.38\times 10^{14}\,$cm$^{-3}$ and $\sigma_{r0}/\delta = 2.9$.
}
\label{fig:5}
\end{figure}
% %%%%%%%%%%%%%%%%%%%%%%%%
\par Figures~\ref{fig:5}(a-d) show maps of the magnitude of the magnetic field calculated from Figs.~\ref{fig:2}(a-d).
The magnetic field is maximum positive %
(convention) at the locations of the bunch filaments (red areas), changes sign around them (blue areas) and vanishes on average outside of the bunch, because, by construction, the plasma return current globally shields the magnetic field of the bunch, but does not shield it locally, i.e. at the scale of the size of the filaments ($\sim \delta$). 
The magnetic field at the center of the filaments is not zero because of the fact that the filaments are not perfectly axisymmetric and they are sorrounded by non-axisymmetric filaments, and because of the finite resolution of the diagnostics. 
The amplitude of the magnetic field reaches $\sim 4\,$mT.
We note that this field amplitude is sufficient to bend the trajectory of a 400\,GeV/c proton by $\delta$ over the plasma length to form filaments.
We also note that the maximum transverse magnetic field generated by the low-current ($\sim$50\,A) bunch at the peak of its density ($t=0$, Fig.~\ref{fig:1}(c)) is $\sim30\,$mT. 
Therefore, even these late filaments generate fields with magnitude comparable to that of the incoming bunch in vacuum.

\par
\begin{figure}[h!]
\centering
\includegraphics[width=\columnwidth]{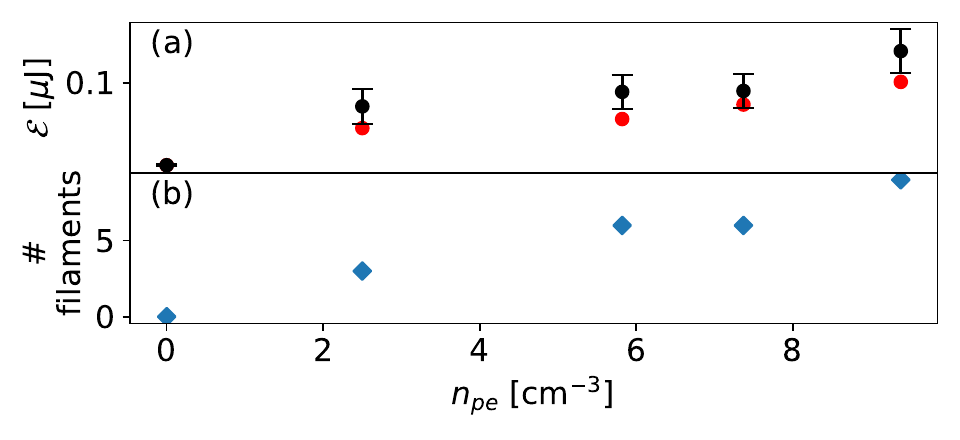}
\caption{
(a): magnetic energy within the bunch volume as a function of $n_{pe}$. 
Black symbols: average of consecutive images; errorbars show the rms variation over the events in each dataset. 
Red symbols: minimum value for each value of $n_{pe}$.
(b): average number of filaments within transverse, time-integrated images at screen B.
Measurements show variation of $\pm1$.
For $n_{pe} = 2.25\times10^{14}\,$cm$^{-3}$, only events showing multiple filaments (see Fig.~\ref{fig:4}) are considered.
}
\label{fig:6}
\end{figure}

 We calculate the amount of magnetic energy within the bunch as $\mathcal{E} = \int dV<B^2>/2\mu_0$, where $V=(2\pi)^{3/2}\sigma_r^2 c\sigma_t$ is the bunch volume with $\sigma_r$ the transverse size at screen B.
Figure~\ref{fig:6}(a) shows that the average magnetic energy (black symbols) increases with $n_{pe}$ and reaches $\sim 0.13\,$\textmu J for $n_{pe} = 9.38\times 10^{14}\,$cm$^{-3}$.
The error bars representing the rms variation over the $>20\,$events at each density are relatively wide. 
This is consistent with the instability nature of the process, producing different number of filaments with different current densities and distributions for every event (see Supplemental Material~\cite{suppl}). 
Since filaments are produced for every event, the minimum energy (red symbols) at each density also increases with $n_{pe}$.
The increase of the magnetic energy with $n_{pe}$ is due to the increase of the maximum amplitude of the magnetic field and to the increase in number of filaments developing within the bunch (Fig.~\ref{fig:6}(b)).
This is consistent with the radius over which filaments are observed in time-integrated images ($r\sim0.5\,$mm in Fig.~\ref{fig:2} and $r\sim0.25\,$mm in Fig.~\ref{fig:4}(c)), and with the transverse modulation depth.
%%%%%%%%%%%%%%%%%%%
% %%%%%%%%%%%%%%%%%%%%%%%%
\par The small (compared to the total kinetic energy of the bunch~$\sim17$\,kJ) amount of magnetic energy shown in Fig.~\ref{fig:6}(a) is consistent with the moderate growth rate (see Eq.~\ref{eq:1}), that causes the instability to occur only late along the bunch with finite duration (Fig.~\ref{fig:3}), with the short length of the interaction and with the low current and current density of the incoming bunch.
Along the earlier part of the bunch, the two currents compensate for each other, and both the amplitude of the magnetic field and the corresponding magnetic energy are close to zero.

\section{Conclusions}
The experimental results presented here show that a long, underdense, relativistic $p^+$ bunch propagating in plasma undergoes the oblique instability when its transverse size is larger than $\sim1.5\,$plasma skin depth.
This is evidenced by the formation of high%
er-current-density filaments, transversely modulating the bunch density and current density.
This can be compared to previous experimental results where the measured threshold for CFI was $\sigma_{r0}\sim2.2$~\cite{CFI:PhysRevLett.109.185007}.
They both confirm the expectation of filamentation to develop when $\sigma_{r0}/\delta>1$.
We also observe the spatio-temporal nature of OBI, manifesting itself through its development only late along the bunch in this case.
Conversely, and most importantly for future plasma wakefield accelerators, filamentation does not occur when the transverse size is smaller than the plasma skin depth. 
Instead, and as measured before, the long $p^+$ bunch undergoes SMI.
At threshold and with the parameters of this experiment, the bunch-plasma system alternates between the two instabilities.
Results also show that filamentation generates magnetic fields in the system, and that the amount of energy converted into magnetic energy increases when increasing the plasma electron density.
If the current filamentation or oblique instability of charged 
particle streams occurred in the universe, they would thus generate the initial magnetic fields that may then be amplified by dynamo processes~\cite{BRANDENBURG20051}.

\begin{acknowledgments}
P. Muggli, L. Verra and E. Walter thank A. Bret for fruitful discussions. 
This work was supported in parts by Fundação para a Ciência e Tecnologia - Portugal (Nos.\ CERN/FIS-TEC/0017/2019, CERN/FIS-TEC/0034/2021, UIBD/50021/2020), STFC (AWAKE-UK, Cockcroft Institute core, John Adams Institute core, and UCL consolidated grants), United Kingdom, the National Research Foundation of Korea (Nos.\ NRF-2016R1A5A1013277 and NRF-2020R1A2C1010835).
M. W. acknowledges the support of DESY, Hamburg.
Support of the Wigner Datacenter Cloud facility through the Awakelaser project is acknowledged.
TRIUMF contribution is supported by NSERC of Canada.
UW Madison acknowledges support by NSF award PHY-1903316.
The AWAKE collaboration acknowledges the SPS team for their excellent proton delivery.
\end{acknowledgments}

\bibliography{second_version}% Produces the bibliography via BibTeX.
\end{document}